\documentclass[11pt]{article}

\usepackage{osid}

\usepackage{floatflt}
\usepackage{epsfig}
\usepackage{amsmath}
\usepackage{amstext}
\usepackage{latexsym}
\usepackage{graphicx}
\usepackage{amsfonts}
\usepackage{caption2}
\usepackage{amsbsy}
\usepackage{amsfonts}
\usepackage[english]{babel}
\usepackage{boxedminipage}
\usepackage{varioref}

\newcommand{\abs}[1]{\left\vert#1\right\vert}

\title{The role of Surface Plasmon modes in the Casimir Effect}
\author{F. Intravaia\thanks{Supported by QUDEDIS, a scientific programme of the
European Science Foundation (ESF).} \\
{\footnotesize\it Universit\"{a}t Potsdam - Institut f\"{u}r Physik,
Am Neuen Palais 10, 14469 Potsdam - Germany \& E-mail:
francesco.intravaia@quantum.physik.uni-potsdam.de}\\[2ex]
A. Lambrecht  \\{\footnotesize\it Laboratoire Kastler-Brossel
ENS,UPMC,CNRS, Universit\'e Pierre et Marie Curie, 4, place Jussieu,
Case 74, F75252 Paris Cedex 05 \& E-mail:
lambrecht@spectro.jussieu.fr} }

\begin{document}

\maketitle
\begin{abstract}
In this paper we study the role of surface plasmon modes in the
Casimir effect. First we write the Casimir energy as a sum over the
modes of a real cavity. We may identify two sorts of modes, two
evanescent surface plasmon modes and propagative modes. As one of
the surface plasmon modes becomes propagative for some choice of
parameters we adopt an adiabatic mode definition where we follow
this mode into the propagative sector and count it together with the
surface plasmon contribution, calling this contribution "plasmonic".
The remaining modes are propagative cavity modes, which we call
"photonic". The Casimir energy contains two main contributions, one
coming from the plasmonic, the other from the photonic modes.
Surprisingly we find that the plasmonic contribution to the Casimir
energy becomes repulsive for intermediate and large mirror
separations. Alternatively,  we discuss the common surface plasmon
defintion, which includes only evanescent waves, where this effect
is not found. We show that, in contrast to an intuitive expectation,
for both definitions the Casimir energy is the sum of two very large
contributions which nearly cancel each other. The contribution of
surface plasmons to the Casimir energy plays a fundamental role not
only at short but also at large distances.
\end{abstract}

\section{Introduction}

An important prediction of quantum theory is the existence of irreducible
fluctuations of electromagnetic fields even in vacuum, that is in the
thermodynamical equilibrium state with a zero temperature. These
fluctuations have a number of observable consequences in microscopic physics
for example in atomic physics the Van der Waals force between atoms in
vacuum.

\begin{figure}[htb] \center
\parbox{7cm}{\epsfig{file=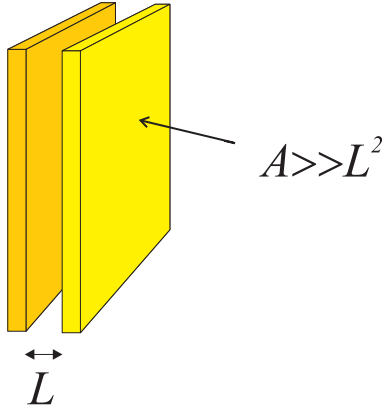,width=5cm}}
\parbox{4.5cm}{\caption{The configuration studied by H. Casimir}
\label{Casimir}}
\end{figure}

Vacuum fluctuations also have observable mechanical effects in
macroscopic physics and the archetype of these effects is the
Casimir force between two mirrors at rest in vacuum. This force was
predicted by H. Casimir in 1948 \cite{Casimir48}  who considered two
plane parallel perfect reflectors as shwon in Figure \ref{Casimir}
%------------------Casimir----------------
%--------------------end figure---------------------
and found an interaction energy $E_{\text{Cas}}$ depending only on
geometrical parameters, the mirrors distance $L$ and surface $A\gg
L^{2}$, and two fundamental constants, the speed of light $c$ and
Planck constant $\hbar $
\begin{equation}
E_{\text{Cas}}=-\frac{\hbar c\pi^2 A}{720L^{3}}.
\label{ECas}
\end{equation}%
The signs have been chosen to fit the thermodynamical convention with
the minus sign of the energy $E_{\text{Cas}}$
corresponding to a binding energy. The Casimir energy for perfect mirrors is usually
obtained by summing the zero-point energies
$\frac{\hbar \omega }{2}$ of the cavity eigenmodes, substracting the
result  for finite and infinite separation, and
extracting the regular expression (\ref{ECas}) by inserting a formal
high-energy cutoff and using the Euler-McLaurin formula \cite{qft}.

The Casimir force was soon observed in different experiments which
confirmed its existence
\cite{Sparnaay89,Milonni94,LamoreauxResource99}. Recent experiments
have reached a good precision, in the \% range, which makes possible
an accurate comparison between theoretical predictions and
experimental observations \cite{Bordag01,Lambrecht02}.

Casimir considered an ideal configuration with two perfectly reflecting
mirrors in vacuum. But the experiments are performed with real reflectors,
for example metallic mirrors which have a perfect reflection only at
frequencies below a plasma frequency $\omega_\mathrm{P}$ or alternatively for mirror separations much larger than the plasma wavelength $\lambda_\mathrm{P}$ characetristic for the metal. Accounting for this imperfect
reflection and its frequency dependence is thus essential for obtaining a
reliable theoretical expectation of the Casimir force in a real situation.

The consideration of real mirrors is important not only for the analysis of
experiments but also from a conceptual point of view. Real mirrors are
certainly transparent at the limit of high frequencies and this allows one
to dispose of the divergences associated with the infiniteness of vacuum
energy. This point was already alluded to in Casimir's papers and an
important step in this direction was the Lifshitz theory of the Casimir
force between two dielectric bulks \cite{Lifshitz56,Schwinger78}. However
this idea was fully implemented in theoretical derivations after a quite
long period.

In the limit of small separations $L \ll\lambda _{\mathrm{P}}$, the Casimir effect has
another interpretation establishing a bridge between quantum field theory of vacuum
fluctuations
and condensed matter theory of forces between two metallic bulks. It can
indeed be understood as resulting from the Coulomb interaction between
surface plasmons, that is the collective electron excitations propagating
on the interface between each bulk and the intracavity vacuum
\cite{Barton79,Schram73}. The corresponding field modes are evanescent waves and
have an imaginary longitudinal wavevector.
At short distances, surface plasmon modes are known to dominate the interaction and
the Casimir energy reduces to \cite{GenetAFLB03,Henkel03}
\begin{equation}
\label{shortapprox} E\approx
E_{\mathrm{pl}}=\frac{3}{2}\alpha\frac{L}{\lambda_{\mathrm{P}}}\quad\text{with
$\alpha=1.193...$ }
\end{equation}

Surface plasmons play an important role in many fields of physics.
Let us only mention as recent examples the surface plasmon assisted
enhancement of the transmission of light through metallic structures
\cite{Ebbesen1,Ebbesen2,Woerdman} or in the context of biomolecular
physics, the importance of plasmon fluctuations of a 2-dimensional
Wigner-like crystal where they can generate an attractive component
to the dispersion forces between parallel surfaces \cite{Pincus}.

%we would appreciate if you would kindly include in the
%Introduction a brief paragraph explaining explicitly the relation of the
%present paper and [20]. 

Here, we derive and expand in more detail the results previously given in Ref. \cite{intravaia:110404},   investigating more closely the influence of surface
plasmons on the Casimir energy, not only at short but at arbitrary
distances. 

\section{Casimir energy for real mirrors}

We restrict our attention to the situation of two infinitely large
plane mirrors at zero temperature so that the only modification of
the Casimir formula (\ref{ECas}) is due to the metals finite
conductivity. This modification is calculated by evaluating the
radiation pressure of vacuum fields upon the two mirrors
\cite{GenetPRA03}
\begin{eqnarray}
E &=&-\sum_{\epsilon}\sum_{\mathbf{k}}\sum_{\omega }\frac{i\hbar}{2}
\ln({1-r_{\mathbf{k}}^{p}[\omega
]^{2}e^{2ik_{z}L}}) + c.c.   \label{Ereal} \\
\sum_{\mathbf{k}} &\equiv &A\int \frac{\mathrm{d}^{2}\mathbf{k}}{4\pi ^{2}}%
\quad ,\quad \sum_{\omega }\equiv \int_{0}^{\infty }\frac{\mathrm{d}\omega }{%
2\pi }\nonumber
\end{eqnarray}%
The energy $E$ is obtained by summing over polarization $p$=(TE,TM),
transverse wavevector $\mathbf{k}\equiv \left( k_{x},k_{y}\right) $
(with $z$ the longitudinal axis of the cavity) and frequency $\omega
$; $k_{z}$ is the longitudinal wavevector associated with the mode.
$r_{\mathbf{k}}^{p}$ are the reflection amplitudes here supposed to
be the same for the two mirrors.

Imperfectly reflecting mirrors will be described by scattering
amplitudes which depend on the frequency, wavevector and
polarization while obeying general properties of stability,
high-frequency transparency and causality. The two mirrors form a
Fabry-Perot cavity with the consequences well-known in classical or
quantum optics~: the energy density of the intracavity field is
increased for the resonant frequency components whereas it is
decreased for the non resonant ones. The Casimir force is but the
result of the balance between the radiation pressure of the resonant
and non resonant modes which push the mirrors respectively towards
the outer and inner sides of the cavity \cite{Jaekel91}. This
balance includes not only the contributions of ordinary waves
propagating freely outside the cavity but also that of evanescent
waves. These two sectors of ordinary and evanescent waves are
directly connected by analyticity properties of the scattering
amplitudes.

Expression (\ref{Ereal}) holds for dissipative mirrors as well as
for non dissipative ones \cite{GenetPRA03}. It
tends towards the ideal Casimir formula (\ref{ECas}) as soon as
the mirrors are nearly perfect for the modes contributing to the
integral.

The reduction of the Casimir energy (\ref{Ereal}) with respect to the
ideal formula (\ref{ECas}) due to the imperfect reflection of
mirrors is described by a factor
\begin{equation}
\eta _{\rm E} = \frac{E}{E_{\rm Cas}}  \label{etaE}
\end{equation}
This factor plays an important role in the discussion of the most
precise recent experiments.

\section{Mode decomposition of the Casimir energy}

We now recalculate the Casimir energy as a sum over
the cavity modes using the plasma model for the mirrors
dielectric function.
\begin{equation}
\varepsilon  (\omega) = 1 - \frac{\omega_{\rm
P}^2}{\omega ^2} \qquad ,\qquad \lambda _{\rm P}=\frac{2\pi c}{\omega
_{\rm P}}
\end{equation}
with $\omega_{\rm P}$ the plasma frequency and $\lambda_{\rm P}$ the
plasma wavelength.

In this case the zeros of the argument of the integrand in
(\ref{Ereal}) lie on the real axis. In fact, they have to be pushed
slightly below this axis by introducing a vanishing dissipation
parameter in order to avoid any ambiguity in expression
(\ref{Ereal}) \cite{GenetPRA03}. We may then rewrite (\ref{Ereal})
as a sum over the solutions $\left[ \omega _{\mathbf{k}}^{p}\right]
_{m}$ of the equation labeled by an integer index $m$
\begin{equation}
r_{\mathbf{k}}^{p}[\omega ]^{2}e^{2ik_{z}L}=1.
\label{solModes}
\end{equation}
Simple algebraic manipulations exploiting residues theorem and
complex integration techniques \cite{Schram73} then lead to the
Casimir energy expressed as sums over these modes
\begin{eqnarray}
E &=& \sum_{p,\mathbf{k}}\left[\sum_{m}^{\prime}\frac{\hbar
\left[ \omega
_{\mathbf{k}}^{p}\right]_{m}}{2}\right]_{L\rightarrow\infty}^{L}.
\label{diff}
\end{eqnarray}%
The prime in the sum over $m$ signifies as usually that the term
$m=0$ has to be multiplied by 1/2. The sum over the modes is to be
understood as a regularized quantity as it involves infinite
quantities. The upper expression contains as limiting cases at
large distances the Casimir expression with perfect mirrors and at
short distances the expression in terms of surface plasmon
resonances (\ref{shortapprox}).

\begin{figure}[htb] \center
\parbox{7cm}{\epsfig{file=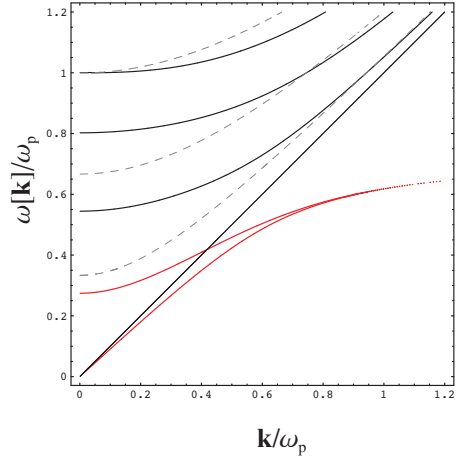,width=6cm}}
\parbox{4.5cm}{\caption{A representation of the dispersion relation
for the TM-modes with the plasma model for
$L/\lambda_{\mathrm{P}}=1.5$. The dashed curves represent the
perfect cavity modes. The solid grey curves represent the plasmonic
modes and the black curves the photonic modes. The $\omega_+$-mode
changes its nature when crossing the light-dispersion curve.}
\label{disp}}
\end{figure}

The ensemble of modes appearing in Eqn. (\ref{diff}) can be
separated into two different ensembles. The TM polarization admits
propagating cavity modes as well as evanescent modes, while the TE
polarization allows only propagating cavity modes. The first
ensemble contains two modes $\omega_+$ and $\omega_-$ which tend to
the usual surface plasmon modes at short distances. $\omega_-$ is
always in the evanescent sector, while $\omega_+$ lies either in the
evanescent or in the propagating sector depending on its parameters
as shown in Fig. \ref{disp}. The second ensemble are propagating
cavity modes which may have TE or TM polarization. In view of the
particular behavior of the $\omega_+$ mode, we adopt an adiabatic
mode definition, where we follow it continuously from the evanescent
into the propagative sector and attribute the whole mode to the
surface plasmon modes, which we call "plasmonic modes"
\cite{intravaia:110404}. This denomination is chosen in order to
avoid confusion with the (common) definition of surface plasmon
modes which defines them as evanescent modes only and cuts the
$\omega_+$ mode into two pieces. Plasmonic modes are thus the cavity
modes living in the evanescent sector at least for some particular
value of their parameters. In the same line of reasoning we call all
propagative modes minus the propagative part of the $\omega_+$ mode
"photonic" modes. Photonic modes are the cavity modes propagating
for all cavity length. In the limit $L\rightarrow\infty$, all
propagative modes tend asymptotically to the eigenmodes of the
perfect cavity. The frequencies of the plasmonic modes $\omega_{+}$
and $\omega_{-}$ degenerate in the limit $L\rightarrow \infty$ to
the surface plasmon frequency $\omega_0$ for a single interface
\cite{heinrichs}
\begin{equation}
\omega_{\pm}\xrightarrow{L\rightarrow
\infty}\omega_{0}[\mathbf{k}]=\sqrt{\frac{\omega^2_P+2\abs{\mathbf{k}}^{2}-\sqrt{\omega^4_P+4\abs{\mathbf{k}}^4}}{2}}
\end{equation}

%------------------disp----------------
%--------------------end figure---------------------

The Casimir energy can now be rewritten as
\begin{equation}
\label{start} E=\sum_{p,\mathbf{k}}\left[\sum_n
\frac{\hbar\omega^p_n}{2}\right]^L_{L\rightarrow\infty}
=\underbrace{ \sum_{\mathbf{k}}\left[\frac{\hbar\omega_+}{2}
+\frac{\hbar\omega_-}{2}\right]^L_{L\rightarrow\infty}}_{\text{plasmonic
Contribution ($E_{\mathrm{pl}}$)}}
+\underbrace{\sum_{p,\mathbf{k}}\left[\sum_{n}
\frac{\hbar\omega^p_n}{2}\right]^L_{L\rightarrow\infty}}_{\text{photonic
Contribution ($E_{\mathrm{ph}}$)}}
\end{equation}
Note that both contributions have no physical meaning on their own, i.e. one cannot measure them separately. The only observable is the total Casimir energy which is the sum of both contributions.
We may rewrite the plasmonic contribution to the Casimir energy
in a more explicit way as follows
\begin{equation}
\label{contr}
E_{\mathrm{pl}}=cA\int\frac{d^2\mathbf{k}}{(2\pi)^2}\frac{\hbar}{2}\left(\omega_+[\mathbf{k},L]+
\omega_-[\mathbf{k},L]-2\omega_{0}[\mathbf{k}]\right)
\end{equation}

The basic idea of the explicit calculation of the plasmonic
contribution to the Casimir energy  resides in the fact that the
frequencies functions $\omega_i,\ i=0,\ \pm$ are solutions of simple
equations. We rewrite Eq.\eqref{contr} in terms of dimensionless
variables
\begin{eqnarray}
\label{defplascontr} E_\mathrm{pl}=\eta_\mathrm{pl}
E_\mathrm{Cas},\quad \eta_\mathrm{pl}=
-\frac{180}{\pi^3}\int_0^{\infty} \sum_i c_i k \
\Omega_i[k]d\mathbf{k}\\
\Omega=\omega L, \quad \Omega_{\mathrm{P}}=\omega_{\mathrm{P}} L,
\quad \abs{\mathbf{k}}L=k, \quad z=k^2-\Omega^2
\end{eqnarray}
with $c_+=c_-=1,c_0=-2$. $\Omega_0$ is the dimensionless surface
plasmon frequency for a single mirror and we have introduced the
corrective factor $\eta_{\mathrm{pl}}$ for the plasmonic
contribution to the Casimir energy. Note that
\begin{equation}
\eta_i=-\frac{180}{\pi^3}\int_0^{\infty} k\ \Omega_i[k]dk,\quad
i=\pm,0
\end{equation}
is divergent despite the convergence of the whole expression  given
in Eq. \eqref{contr}. Without giving any details let us just mention
that in order to perform the explicit calculation we need to
introduce a regularizing factor. Such a modification is
mathematically mathematical convenient and does not affect the final
result.

It can be shown that the dimensionless frequencies $\Omega_i[k]$
can be formally obtained as
\begin{equation}
\label{invert}
k^2=f_i(z)\Rightarrow
\Omega_i[k]=\sqrt{k^2-f_i^{-1}[k^2]},\quad i=0,\pm
\end{equation}
where
\begin{subequations}
\begin{gather}
f_+(z)=z+\frac{\Omega^2_p\
\sqrt{z}}
{\sqrt{z}+\sqrt{z+\Omega_{\mathrm{P}}^2}\tanh[\frac{\sqrt{z}}{2}]}=z+g_+^2[z]\\
f_-(z)=z+\frac{\Omega^2_p\ \sqrt{z}}
{\sqrt{z}+\sqrt{z+\Omega_{\mathrm{P}}^2}\coth[\frac{\sqrt{z}}{2}]}=z+g_-^2[z]\\
f_0(z)=z+\frac{\Omega^2_p\sqrt{z}}
{\sqrt{z}+\sqrt{z+\Omega_{\mathrm{P}}^2}}=z+g_0^2[z]
\end{gather}
\end{subequations}
Let us also define
\begin{equation}
y^{2}_{i}=z^{0}_{i}=-f_i^{-1}[0]=\Omega^{2}_i[0]
\end{equation}
With the change of variable $k^2=z+g^2_i(z)$ and after some
rearrangements Eqn. (\ref{defplascontr}) can be rewritten as
\begin{equation}
\label{fin}
\eta_{pl}=-\frac{180}{2\pi^3}\left[
\int_0^{\infty}\sum_i c_i
g_i(z)dz+\int_{-z_+^0}^{0}g_+(z)dz-\frac{2}{3}y^3_+\right]
\end{equation}
where we
have exploited the fact that
\begin{gather}
z^{0}_{+}\not=z^{0}_{-}=z^{0}_{0}=0, \quad f_i^{-1}[\infty]=\infty
\nonumber \\ f_i(-z_i^0)=0\Rightarrow g_i(-z_i^0)=\sqrt{z_i^0} \quad
\text{and} \quad g_i(\infty)=\frac{\Omega_{\mathrm{P}}}{\sqrt{2}}\nonumber
\end{gather}
The corrective factor $\eta_\mathrm{pl}$ has a well defined
structure: it is indeed decomposed into an integral over the
positive real $z$-axis plus an integral over an interval of the
negative $z$-axis plus a constant depending only on
$\Omega_{\mathrm{P}}$. Moreover only $g_+$, is involved in the last
integral and in the constant. This particular structure can be
traced back to the properties of the plasmonic modes. The positive
$z$-value domain coincides with the evanescent sector while the
negative one describes the propagative sector. While the plasmonic
mode $\omega_-$ and the surface plasmon frequency for a single
mirror $\omega_{0}$ are totally contained in the evanescent sector,
the plasmonic mode $\omega_+$ lives in both sectors. Therefore $g_0$
and $g_-$ describing the properties of $\omega_-$ and $\omega_{0}$
are contained only in the first integral while $g_+$ has to be
evaluated in a wider range of $z$-values which includes an interval
in the propagative sector. The second integral in Eq. \eqref{fin} is
thus basically the propagative part contribution of the plasmonic
mode $\omega_+$.

\begin{figure}
\center\epsfig{file=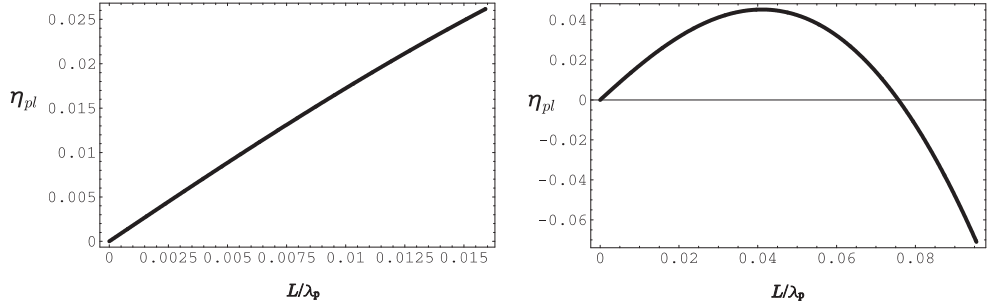,width=13cm} \caption{The normalized
plasmonic mode contribution to the Casimir force as function of
$L/\lambda_{\mathrm{P}}$ for two different distance intervals.}
\label{pat}
\end{figure}

Fig. \ref{pat} shows the numerical evaluation of Eqn. (\ref{fin})
for $\eta_\mathrm{pl}$ as function of $L/\lambda_{\mathrm{P}}$ for
two different distance intervals. The left graphic illustrates the
short distance behavior In the limit $\Omega_\mathrm{P}\ll 1$ the
corrective factor $\eta_\mathrm{pl}$ can be approximated by the
first integral of Eq. \eqref{fin}.
%----------------------graphics eta---------------------------
%---------------end figure-----------------------------
leading to
\begin{equation}
\label{s}
\eta_{\mathrm{pl}}\approx-\frac{180}{2\pi^3} \frac{\Omega_{\mathrm{P}}}{\sqrt{2}}
\int_{0}^{\infty}\left(\sqrt{1+e^{-\sqrt{z}}}+\sqrt{1-e^{-\sqrt{z}}}-2\right)dz=\frac{3}{2}\alpha\frac{L}{\lambda_{\mathrm{P}}}
\end{equation}
We recover here the result of Eqn. (\ref{shortapprox}). The right
graphic in Fig. \ref{pat} shows the plasmonic mode contribution
$\eta_{\mathrm{pl}}$ at large distances. Surprisingly, it changes
its sign for $\frac{L}{\lambda_{\mathrm{P}}}\sim 0.08$ and its slope
and diverges for $L\gg\lambda_{\mathrm{P}}$ \cite{intravaia:110404}.
In the large distance limit $\Omega_{\mathrm{P}}\gg 1$ we find the
following asymptotic behavior
\begin{equation}
\label{ldae}
\eta_{\mathrm{pl}}\approx-\Gamma\sqrt{\Omega_{\mathrm{P}}}, \qquad \Gamma=29.752
\end{equation}
The contribution of the plasmonic modes to the Casimir energy becomes thus repulsive for intermediate and large distances. This result is based on an adiabatic definition of the surface plasmon modes, where we follow the $\omega_+$ mode even when it crosses the barrier $\omega=ck$ and becomes propagative.

Let us now compare this result to the common definition of surface
plasmon modes which includes only evanescent waves and cuts the
$\omega_+$ mode at $\omega=ck$ as for example done by Bordag
recently \cite{Bordag2005}. This leads to
\begin{equation}
\eta_\mathrm{ev}=-\frac{180}{\pi^{3}}\int_{k_{\mathrm{P}}}^{\infty}k\left(
\Omega_{+}[k]-\Omega_{0}[k]\right)dk-\frac{180}{\pi^{3}}\int_{0}^{\infty}k\left(\Omega_{-}[k]-\Omega_{0}[k]\right)dk
\end{equation} where
\begin{equation}
k_{\mathrm{P}}=g_{+}(0)=\Omega_{\mathrm{P}}/\sqrt{1+\Omega_{\mathrm{P}}/2}
\end{equation}
is associated with the value of $\abs{\mathbf{k}}$ for which this
modes crosses the light cone. $\eta_\mathrm{ev}$ can be evaluated
numerically \cite{Bordag2005}, or using the same method developed
here
\begin{equation}
\label{etab}
\eta_\mathrm{ev}=-\frac{180}{2\pi^3}\left(
\int_0^{\infty}\sum_i c_i g_i(z)dz-\int_{-z^{P}_{0}}^{0}
g_0(z)dz-\frac{2}{3}\left(k^{3}_{\mathrm{P}}-\Omega^{3}_{0}[k_{\mathrm{P}}]
\right)\right)
\end{equation}
where we have exploited the following relations
\begin{equation}
-z^{P}_{0}=f^{-1}_{0}[k_{\mathrm{P}}]=k^{2}_{\mathrm{P}}-\Omega^{2}_{0}[k_{\mathrm{P}}]\Rightarrow
g_{0}(-z^{P}_{0})=\Omega_{0}[k_{\mathrm{P}}]
\end{equation}
Eq. \eqref{etab} shows the short distance asymptotic behavior of the
total Casimir energy given in \eqref{shortapprox} as it was reported
in \cite{GenetAFLB03,Henkel03}. It does not change  its sign and, in
agreement with \cite{Bordag2005}, at long distances goes as
\begin{equation}
\eta_{\mathrm{ev}}=\beta_\mathrm{ev}\sqrt{\Omega_{\mathrm{P}}}\quad\text{with}\quad
\beta_\mathrm{ev}=1.62399...
\end{equation}
With this definition we  therefore naturally find the result that
the contribution of the evanescent modes to the Casimir energy is
always attractive and reproduces well the short distance behavior of
the Casimir energy.

\section{Conclusion}

As a concluding remark we would like to stress that, no matter how
we attribute the propagative part of the $\omega_+$ mode, whether to
the surface plasmon modes in an adiabatic definition (plasmonic
modes) or to the propagating modes, the influence of surface
plasmons is very important at all distances. The Casimir energy is
the detailed balance of two very large contributions of opposite
sign which nearly cancel each other, but not quite, the difference
being a small Casimir energy, much smaller than each of both
contributions.

It might be interesting to investigate if a change in the
photon-plasmon coupling could somehow influence this detailed
balance and therefore the value or even the sign of the Casimir
force. A different coupling could be obtained by using for example
nanostructured surfaces. For such an analysis the adiabatic mode
definition should be well suited, because for a different coupling
the mode will change \textit{as a whole} and in following
continuously the mode one could trace back the changes introduced
through the structured surface.

\end{document}